\documentclass[letterpaper]{article} 
\usepackage{aaai2026}  
\usepackage{times}  
\usepackage{helvet}  
\usepackage{courier}  
\usepackage[hyphens]{url}  
\usepackage{graphicx} 
\usepackage{natbib}  
\usepackage{caption} 
\usepackage{algorithm}
\usepackage{algorithmic}
\usepackage{booktabs}
\usepackage{multirow}
\usepackage{amsmath}
\usepackage{tikz}
\usepackage{pgfplots}
\pgfplotsset{compat=1.18}
\usepackage{amssymb}
\usepackage{hyperref}
\usepackage{newfloat}
\usepackage{listings}
\DeclareCaptionStyle{ruled}{labelfont=normalfont,labelsep=colon,strut=off} 
\floatstyle{ruled}
\newfloat{listing}{tb}{lst}{}
\floatname{listing}{Listing}
\title{New Synthetic Goldmine: Hand Joint Angle-Driven EMG Data Generation Framework for Micro-Gesture Recognition}
\author{
    Nana Wang\textsuperscript{\rm 1,2}, 
    Gen Li\textsuperscript{\rm 1},     
    Pengfei Ren\textsuperscript{\rm 3},
    Hao Su\textsuperscript{\rm 4},
    Suli Wang\textsuperscript{\rm 5}\thanks{Corresponding author: \href{mailto:wsuli615@gmail.com}{wsuli615@gmail.com}.}
}
\affiliations{
    \textsuperscript{\rm 1}School of Computer Science and Engineering, Beihang University, China\\
    \textsuperscript{\rm 2}NoBarriers.ai Technology, Hangzhou, China\\
    \textsuperscript{\rm 3}Beijing University of Posts and Telecommunications, China\\
    \textsuperscript{\rm 4}Zhengzhou University, China\\
    \textsuperscript{\rm 5}Department of Computer Science, Technische Universität Darmstadt, Germany
}

\usepackage{bibentry}

\begin{document}

\maketitle

\begin{abstract}
Electromyography (EMG)-based gesture recognition has emerged as a promising approach for human-computer interaction. However, its performance is often limited by the scarcity of labeled EMG data, significant cross-user variability, and poor generalization to unseen gestures. To address these challenges, we propose SeqEMG-GAN, a conditional, sequence-driven generative framework that synthesizes high-fidelity EMG signals from hand joint angle sequences. Our method introduces a context-aware architecture composed of an angle encoder, a dual-layer context encoder featuring the novel Ang2Gist unit, a deep convolutional EMG generator, and a discriminator, all jointly optimized via adversarial learning. By conditioning on joint kinematic trajectories, SeqEMG-GAN is capable of generating semantically consistent EMG sequences, even for previously unseen gestures, thereby enhancing data diversity and physiological plausibility. Experimental results show that classifiers trained solely on synthetic data experience only a slight accuracy drop (from 57.77\% to 55.71\%). In contrast, training with a combination of real and synthetic data significantly improves accuracy to 60.53\%, outperforming real-only training by 2.76\%. These findings demonstrate the effectiveness of our framework,also achieves the state-of-art performance in augmenting EMG datasets and enhancing gesture recognition performance for applications such as neural robotic hand control, AI/AR glasses, and gesture-based virtual gaming systems.
\end{abstract}

\section{Introduction}
Electromyography (EMG)-based gesture recognition has become a key modality for human-computer interaction (HCI), owing to its noninvasive sensing and ability to capture subtle gesture movement. It has shown great promise in applications such as neural prosthetics, AR/VR interfaces, and wearable assistive technologies. However, practical deployment remains constrained by several critical challenges: the scarcity of large-scale and diverse EMG datasets, limited generalization across users and tasks, and the difficulty in modeling temporal and semantic consistency of EMG patterns~\cite{eddy2023framework}.

A major bottleneck lies in the scarcity and structural limitations of existing EMG datasets. Publicly available datasets such as Ninapro~\cite{atzori2012building}, CapgMyo~\cite{dai2021capg}, CSL-HDEMG~\cite{amma2015advancing}, and SEEDS~\cite{matran2019seeds} are predominantly collected under predefined and isolated gesture protocols, offering limited degrees of freedom (DoFs) and failing to support natural transitions or combinations of DoFs commonly required in real-world control scenarios. This lack of flexibility restricts the datasets’ utility for modeling intuitive, user-driven interactions. Although HIT-SimCo~\cite{yang2021emg} attempts to address this issue by recording randomized combinations of wrist DoFs, it is constrained to single-session recordings, thereby limiting its applicability to studies involving cross-day variability, learning adaptation, and long-term stability. The recently introduced \textit{emg2pose} dataset~\cite{salter2024emg2pose} significantly advances the state of EMG data collection by offering large-scale, multi-user, and multi-session recordings with synchronized hand pose annotations. It comprises over 370 hours of data from 193 subjects, covering 29 gesture stages under varied sensor placements and session conditions. Compared to earlier datasets, \textit{emg2pose} provides high-resolution joint-angle trajectories across the entire hand and introduces benchmark splits to evaluate generalization between users, between sessions and between tasks. An additional concern is the significant imbalance between dynamic and isometric contractions, which can lead to biased model performance and reduced generalizability across different contraction types.


On the algorithmic side, effective EMG pattern recognition remains highly dependent on capturing the complex temporal and semantic relationships inherent in muscle signals. While earlier studies employed traditional machine learning with handcrafted features~\cite{{phinyomark2018emg},{oskoei2007myoelectric},{oskoei2008support}}, such methods often suffer from suboptimal feature representations and poor scalability. Although recent deep learning approaches~\cite{gallon2024comparison,xiong2021deep,chen2020surface} have improved recognition performance through end-to-end learning, several critical issues persist. 
Despite recent advances, EMG-based gesture modeling continues to face three interrelated challenges. First, many existing models struggle to capture the intricate temporal dependencies between joint kinematics and EMG activation, resulting in decoded or generated signals that lack semantic coherence and physiological plausibility~\cite{tacca2024wearable,sitole2023continuous,zhang2022ankle}. Second, the scarcity and uneven distribution of gesture data hinder model robustness, while commonly adopted augmentation strategies—such as random noise injection~\cite{fatayer2022semg} or label-conditioned generation~\cite{xiong2024patchemg}—lack the precision and coverage required to synthesize high-quality samples in underrepresented regions of the EMG feature space~\cite{nasrallah2023semg,jiang2021data}. Third, current discriminator-based evaluation frameworks rely on narrow, signal-level criteria and are ill-equipped to assess the fidelity of generated EMG signals across multiple dimensions, including amplitude continuity, temporal alignment, and waveform morphology~\cite{farago2022review,zhao2024trgan,tsinganos2020data}.

These challenges are further amplified in the context of EMG signal generation. Unlike image or audio synthesis, generating realistic and semantically meaningful EMG signals is fundamentally constrained by the physiological complexity, non-stationary dynamics, and user-specific variability of muscle activation. Existing generative models—often based on label-conditioned or noise-driven inputs—fail to provide fine-grained control and struggle to maintain semantic consistency and physiological plausibility, particularly in gesture-sparse regions. Moreover, the lack of comprehensive, physiology-aware evaluation mechanisms exacerbates the difficulty of training reliable EMG generators. These limitations highlight the need for a controllable, physiologically grounded EMG generation framework that ensures both semantic consistency and distributional completeness.

Joint angle–driven conditioning is a promising direction: joint angles summarize underlying activations, and conditioning on them renders EMG synthesis more plausible under inverse-dynamics constraints on feasible muscle activity. Here we introduce SeqEMG-GAN, a sequence-driven conditional generative model that synthesizes high-fidelity EMG signals based on hand joint kinematic trajectories. Our key motivation is to leverage joint angle sequences—an interpretable and compact representation of movement—as a prior control signal to guide the generation of temporally aligned, physiologically consistent EMG sequences. The proposed framework comprises four core components: (1) a joint angle encoder to extract movement embeddings, (2) a latent context encoder incorporating a novel Ang2Gist unit to model fine-grained temporal dependencies between motion and muscle activation, (3) a deep convolutional generator to produce raw EMG waveforms, and (4) a multi-perspective discriminator that evaluates signal fidelity across amplitude, timing, and morphological perspectives. This design enables both targeted synthesis in data-scarce regions and robust generalization across gestures and users.

The main contributions of this work are as follows:
\begin{itemize}
\item \textbf{A latent context encoder} incorporating a novel \textit{Ang2Gist} unit to capture temporal dependencies between joint movement and EMG activation, ensuring semantic and physiological coherence in generated signals;
\item \textbf{A distribution-aware conditional augmentation strategy} that utilizes joint-angle conditioning to guide sample generation in underrepresented regions of the EMG feature space, improving data diversity and coverage;
\item \textbf{A multi-perspective discriminator} that jointly evaluates the fidelity of synthetic signals across amplitude profiles, temporal alignment, and waveform morphology, enhancing the realism of generated EMG data.
\end{itemize}

Together, these components form a controllable and physiologically grounded EMG generation pipeline, offering an effective solution for data augmentation under limited or imbalanced data regimes, and opening new directions for EMG-based gesture recognition in neural interface applications.

\section{Related Work}

\subsection{Traditional EMG Data Augmentation and Generation Methods}

To augment scarce EMG data, early attempts such as temporal warping~\cite{gu2018robust} and noise injection~\cite{stachaczyk2020adaptive} have been explored, but they lack semantic consistency. Transfer learning~\cite{ameri2019deep, wu2023transfer} alleviates inter-user variability, yet it relies on large-scale source data. Earlier biomechanical models~\cite{kiguchi2008development} map EMG to joint motion via physical equations, but their complexity and reliance on expert knowledge limit scalability. Statistical models like GMM and HMM~\cite{yang2017semg} impose strong assumptions, reducing generative diversity and generalization across poses and users.

These limitations motivate the need for data-driven generative approaches that can capture both temporal dynamics and individual variability.

\begin{figure*}[!ht]
    \centering
    \includegraphics[width=0.95\linewidth]{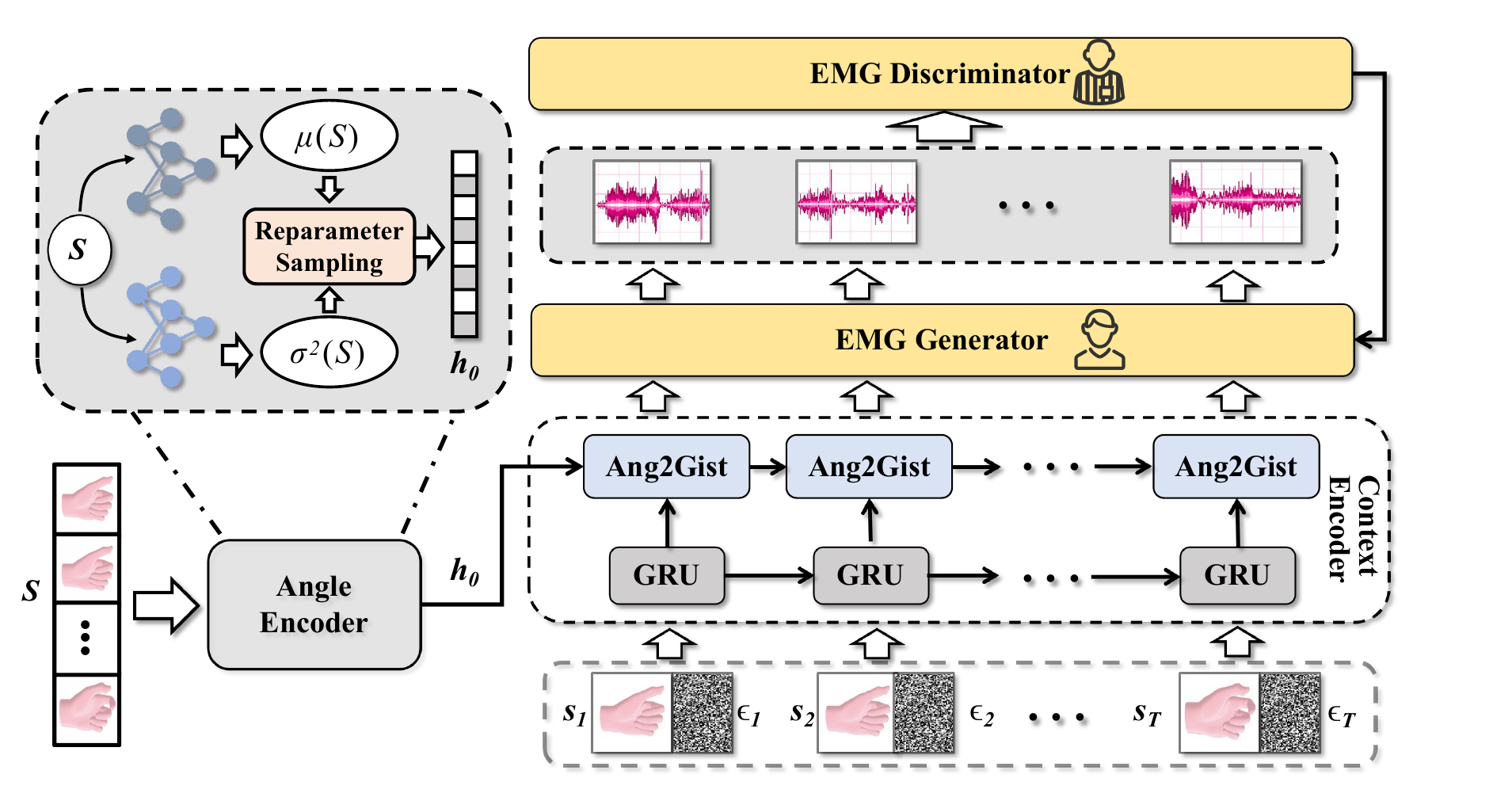}
    \caption{Overview of the SeqEMG-GAN framework. Joint angle sequences are encoded into a global latent variable and contextual hidden states, which guide the generator in synthesizing EMG signals. A discriminator jointly evaluates temporal and semantic alignment.}
    \label{fig:enter-label}
\end{figure*}

\subsection{Generative Methods for EMG}
Generative Adversarial Networks (GANs), Variational Autoencoders (VAEs), and diffusion-based models have emerged as foundational tools to address EMG-specific challenges such as data scarcity, signal degradation, and inter-subject variability.

\textbf{GAN-Based Methods:} \cite{mendez2022emg} applied GANs to generate synthetic EMG signals for hand grasp classification, leading to improved classifier performance in low-resource settings. \cite{chen2022deep} employed DCGANs to augment multi-channel EMG data represented as grayscale images, achieving marginal but consistent gains in recognition accuracy. \cite{lin2023toward} proposed RoHDE, a GAN-based framework that introduces perturbation patterns into HD-EMG to enhance classifier robustness against real-world signal artifacts. \cite{lin2023toward} introduced a style-transfer GAN to disentangle identity-specific signal features from gesture-related information, highlighting privacy vulnerabilities and EMG anonymization opportunities.

\textbf{Privacy-Preserving Synthesis:} Beyond task-driven synthesis, \cite{hazra2020synsiggan} designed SynSigGAN to generate synthetic biomedical signals, including EMG, achieving high signal correlation and enabling data sharing under privacy constraints.

\textbf{Diffusion-Based Methods:} Recent diffusion-based approaches have shown superior modeling of complex EMG distributions. \cite{xiong2024patchemg} introduced PatchEMG, which employs patch-wise diffusion for few-shot gesture recognition, demonstrating high fidelity synthesis with limited data. \cite{liu2024sdemg} proposed SDEMG, a score-based diffusion model that effectively removes ECG artifacts from diaphragmatic EMG. \cite{qing2024enhancing} utilized a diffusion model to synthesize realistic EMG for ergonomic studies, preserving utility while mitigating privacy concerns.

\textbf{VAE-Based Methods:} Latent-variable models offer calibrated uncertainty that is valuable for open-set scenarios. ~\cite{olsson2021can} propose a VAE-based framework for open-set gesture recognition using sparse multichannel EMG signals. Their method enables reliable classification of known gestures while rejecting unknown ones, addressing a key need in real-world prosthetic control. Unlike closed-set models, the VAE captures uncertainty, allowing effective separation of seen and unseen gestures.




Together, these contributions demonstrate the versatility of generative models in advancing EMG-based applications across signal enhancement, data augmentation, privacy, and human-computer interaction.

\section{Methodology}

We propose \textbf{SeqEMG-GAN}, a sequence-driven conditional generative framework that synthesizes electromyographic (EMG) signal sequences from hand joint kinematic inputs. The architecture is designed to model complex temporal dependencies and generate physiologically plausible EMG signals, even for underrepresented gestures. As shown in Figure~\ref{fig:enter-label}, SeqEMG-GAN consists of four main modules: Angle Encoder, Context Encoder, EMG Generator, and EMG Discriminator.

\subsection{Overall Framework}

Given a joint angle sequence $S = \{s_1, s_2, \dots, s_T\}$, the model generates a corresponding EMG sequence $\hat{X} = \{\hat{x}_1, \hat{x}_2, \dots, \hat{x}_T\}$. The generation process is governed by a global latent context vector $\mathbf{h}_0$, which summarizes high-level motion semantics. To enable stochastic generation and improve generalization, we apply the \textit{reparameterization trick}~\cite{kingma2015variational} to sample $\mathbf{h}_0$ from a learned posterior distribution:

\begin{equation}
    \mathbf{h}_0 = \mu(S) + \Sigma(S)^{1/2} \odot \epsilon_s, \quad \epsilon_s \sim \mathcal{N}(\mathbf{0},\mathbf{I}).
\end{equation}

Here, $\mu(S)$ and $\Sigma(S)$ are the outputs of neural networks conditioned on $S$. This stochastic sampling introduces variability into the generated signals while maintaining differentiability during training.

\subsection{Context Encoding via Ang2Gist}

The \textbf{Context Encoder} models sequential dependencies between joint movement and muscle activation. It is implemented as a two-layer recurrent structure:
\begin{itemize}
    \item \textbf{Lower Layer:} Standard Gated Recurrent Unit (GRU) units encode motion dynamics.
    \item \textbf{Upper Layer:} A novel Angle‑to‑Gist (Ang2Gist) unit refines contextual representation.
\end{itemize}

At each timestep $t$, we compute:
\begin{equation}
    \mathbf{i}_t, \mathbf{g}_t = \text{GRU}( \mathbf{s}_t \| \epsilon_t, \mathbf{g}_{t-1}),
\end{equation}
where $\epsilon_t \sim \mathcal{N}(0, I)$ is noise and $\|$ denotes concatenation. The output $\mathbf{i}_t$ is passed to the Ang2Gist unit:

\begin{equation}
    \mathbf{o}_t, \mathbf{h}_t = \text{Ang2Gist}(\mathbf{i}_t, \mathbf{h}_{t-1}),
\end{equation}
producing a Gist vector $\mathbf{o}_t$ for EMG synthesis and updating the hidden state $\mathbf{h}_t$. The Ang2Gist computations are defined as:

\begin{equation}
\begin{aligned}
    \mathbf{z}_t &= \sigma_z(\mathbf{W}_z \mathbf{i}_t + \mathbf{U}_z \mathbf{h}_{t-1} + \mathbf{b}_z), \\
    \mathbf{r}_t &= \sigma_r(\mathbf{W}_r \mathbf{i}_t + \mathbf{U}_r \mathbf{h}_{t-1} + \mathbf{b}_r), \\
    \mathbf{h}_t &= (1 - \mathbf{z}_t) \odot \mathbf{h}_{t-1} + \mathbf{z}_t \odot \sigma_h(\mathbf{W}_h \mathbf{i}_t + \mathbf{U}_h \mathbf{h}_{t-1} + \mathbf{b}_h), \\
    \mathbf{o}_t &= \text{Filter}(\mathbf{i}_t) \ast \mathbf{h}_t.
\end{aligned}
\end{equation}

Here, $\sigma_{\{z,r,h\}}$ are nonlinear activations; $\mathbf{W}_\ast$, $\mathbf{U}_\ast$, and $\mathbf{b}_\ast$ are trainable parameters; $\odot$ and $\ast$ denote element-wise product and filtering, respectively. Filter(·) is a 1D depthwise‑separable convolution along time (kernel size = 5, stride = 1, padding = 2) applied to the angle embedding $i_t$; its parameters are learned jointly with the generator (Xavier‑uniform initialization, SGD with momentum), and each filter channel is shared across gesture dimensions to enforce temporal consistency at low cost.

The \textbf{Ang2Gist} unit is designed to address the challenge of capturing \textit{temporally consistent} and \textit{semantically aligned} representations for EMG signal generation. Standard GRU-based encoders often fail to retain local dynamics while preserving global contextual consistency. Ang2Gist explicitly fuses motion-specific features \(\mathbf{i}_t\) with evolving context states \(\mathbf{h}_t\), enabling the generator to condition on both instantaneous motion and accumulated semantics. This design enhances the generator's ability to produce physiologically plausible and temporally coherent EMG sequences.

\subsection{EMG Generation and Discrimination}

The \textbf{EMG Generator} synthesizes EMG signals at each timestep using the corresponding Gist vector:
\begin{equation}
    \hat{\mathbf{x}}_t = \mathcal{G}(\mathbf{o}_t),
\end{equation}
where $\mathcal{G}$ is implemented as a multi-layer convolutional decoder with transposed convolutions for temporal upsampling.

The \textbf{EMG Discriminator} evaluates whether the generated EMG signal $\hat{\mathbf{x}}_t$ is realistic and semantically aligned with the joint angle sequence $S_t$, conditioned on global context $\mathbf{h}_0$:
\begin{equation}
    m_t = \mathcal{D}(\hat{\mathbf{x}}_t, S_t \mid \mathbf{h}_0).
\end{equation}

\subsection{Implementation}

The proposed \textbf{SeqEMG-GAN} framework generates EMG signal sequences from joint kinematic inputs, maintaining temporal coherence and physiological plausibility. The model is built on a conditional GAN architecture and consists of four core components:
\begin{itemize}
    \item \textbf{Angle Encoder:} Encodes joint angle sequences into compact latent representations.
    \item \textbf{Context Encoder:} Models temporal dependencies using recurrent layers and produces a global context vector $h_0$.
    \item \textbf{EMG Generator ($G$):} Synthesizes EMG signals from joint embeddings, noise, and context.
    \item \textbf{EMG Discriminator ($D$):} Evaluates the realism and motion-alignment of generated EMG sequences.
\end{itemize}

\subsubsection{Data Representation}

The training data consists of paired samples $\left\{(S^{(i)}, X^{(i)})\right\}_{i=1}^N$, where each pair contains:
\begin{itemize}
    \item \textbf{Joint angle sequence} $S \in \mathbb{R}^{T \times d_j}$, representing the hand gesture over $T$ time steps
    \item \textbf{EMG sequence} $X \in \mathbb{R}^{T \times d_e}$, capturing the corresponding muscle activation signals
\end{itemize}

The joint angle sequence $S$ is used as the conditioning input to guide the generation of the EMG signal $X$ through the SeqEMG-GAN model.

\subsection{Training Objective}

The learning objective integrates a standard adversarial loss and a Kullback–Leibler (KL) divergence regularization:

\begin{equation}
\min_{\theta} \max_{\psi} \left( \alpha \, \mathcal{L}_{\text{GAN}} + \mathcal{L}_{\text{KL}} \right)
\end{equation}

where $\theta$ and $\psi$ are the parameters of the generator and discriminator, respectively, and $\alpha$ is a balancing hyperparameter.

\subsubsection{Adversarial Loss}

The adversarial loss encourages the generator to synthesize realistic EMG sequences conditioned on joint angle sequences:

\begin{equation}
\begin{aligned}
\mathcal{L}_{\text{GAN}} = &\; \mathbb{E}_{(x_t, s_t)} \left[ \log D(x_t, s_t, h_0) \right] \\
& + \mathbb{E}_{(\epsilon_t, s_t)} \left[ \log \left( 1 - D(G(\epsilon_t, s_t; \theta), s_t, h_0) \right) \right]
\end{aligned}
\end{equation}

Here, $x_t$ denotes the real EMG signal, $s_t$ is the joint angle input, $\epsilon_t \sim \mathcal{N}(0, I)$ is the noise input, and $h_0$ is the global context variable.

\subsubsection{KL Divergence Regularization}

To ensure a structured latent space, a KL divergence regularization term constrains the global latent variable $h_0$ to follow a standard normal distribution:

\begin{equation}
\mathcal{L}_{\text{KL}} = D_{\text{KL}} \left( \mathcal{N}(\mu(S), \text{diag}(\sigma^2(S))) \, \| \, \mathcal{N}(0, I) \right)
\end{equation}

Here, $\mu(S)$ and $\sigma^2(S)$ are parameterized by neural networks and computed from the joint sequence $S$.

\section{Experiments}

\subsection{Dataset and Experimental Setup}

The experiments in this chapter utilize the latest \textbf{emg2pose} dataset proposed by Meta~\cite{salter2024emg2pose}. This dataset is currently the largest publicly available high-quality dataset for wrist-based electromyography (EMG) recordings with hand gesture annotations. The \textbf{emg2pose} dataset includes 16-channel EMG signals sampled at 2 kHz, along with precise gesture annotations captured by 26 cameras. It covers a total of 193 users, 370 hours of recording time, and 29 gesture categories, ensuring a high degree of gesture diversity.

During the experiments, the dataset is randomly split into training, validation, and test sets in an 8:1:1 ratio. Specifically, the training set contains 20,203 EMG samples, while the validation and test sets each contain 2,525 EMG samples. The scale and precision of this dataset make it an ideal benchmark for EMG-based gesture recognition research, providing a reliable data foundation for the experiments in this chapter.
\begin{figure}[t]
    \centering
    \includegraphics[width=\linewidth]{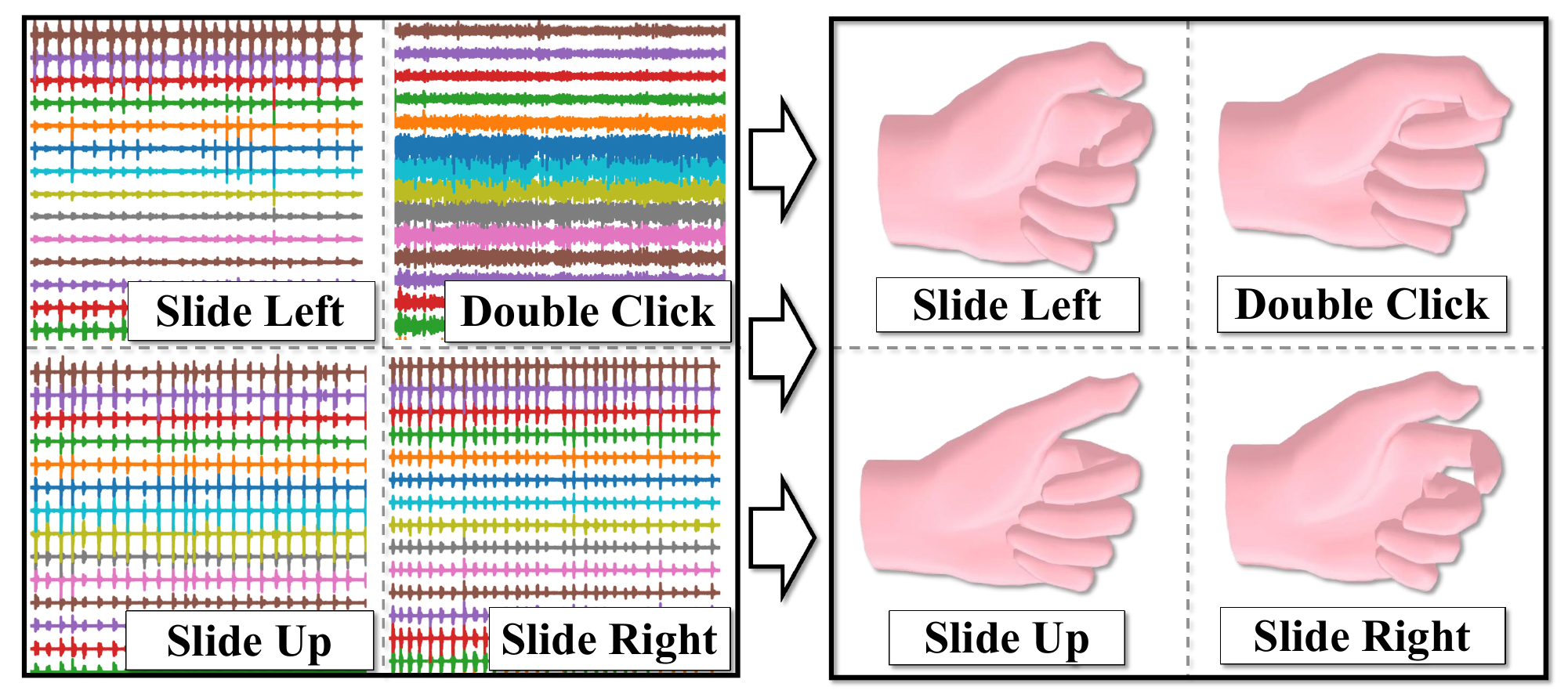}
    \caption{Synthesized EMG Gesture Data and Its Visualization Results by SeqEMG-GAN}
    \label{fig:show}
\end{figure}

\subsection{Implementation Details}

In the experiments, the batch size is set to 32, and training is conducted for 100 epochs. The initial learning rate is set to 0.002, with an annealing strategy applied during the pretraining phase to gradually decay the learning rate. During model fine-tuning, the learning rate is reduced by a factor of 10 at the 21st, 24th, and 27th epochs.

The training process utilizes the Stochastic Gradient Descent (SGD) optimizer, with a momentum hyperparameter of 0.9 and a weight decay of $5\times10^{-4}$. The weight parameter $\lambda$ is set to 2. All experiments are implemented using the PyTorch library and trained on two NVIDIA RTX 3090 Ti GPUs.

\subsection{Evaluation Metrics}

To assess the effectiveness of SeqEMG-GAN, we conducted two main experiments: \textit{similarity analysis} and \textit{performance validation}. Given that standard GAN evaluation metrics such as Inception Score (IS), Frechet Inception Distance (FID), and Kernel MMD~\cite{xu2018diversity} are primarily designed for image generation and high-dimensional feature space comparison, they are not directly applicable to EMG time-series data.

To address this, we adopt metrics more suitable for temporal biomedical signals, following prior work~\cite{delaney2019synthesis}, and propose a multi-perspective evaluation scheme covering frequency-domain, time-domain, and signal morphology consistency:

\paragraph{Fast Fourier Transform MSE (FFT MSE)}
We compute the mean squared error between the FFT magnitudes of generated and real EMG signals. This metric evaluates frequency-domain similarity and reflects how well the generated signals preserve spectral characteristics essential to muscle activation patterns.

\paragraph{Dynamic Time Warping (DTW)}
To measure structural similarity in the time domain, we use DTW to non-linearly align generated and real sequences. DTW is a distance between two time series that allows non-linear time stretching/compression (lower is better). This technique is especially robust to temporal shifts and warping in EMG signals. We implement FastDTW~\cite{salvador2007toward} to reduce computational cost to \( O(N) \).

\paragraph{EMG Envelope Cross-Correlation (CC)}
We evaluate global shape consistency using cross-correlation between smoothed signal envelopes. The envelope is computed by applying a moving average filter to the absolute EMG signals, and normalized CC quantifies trend alignment and tremor consistency.

Together, these metrics provide a comprehensive evaluation framework for EMG signal generation, capturing fidelity across spectral, temporal, and morphological dimensions.

\begin{table*}[!ht]
    \centering
    \resizebox{\textwidth}{!}{
    \begin{tabular}{lccccc}
        \toprule
        \textbf{Model} & \textbf{Conditional Generation} & \textbf{Unknown Gesture Generation} & \textbf{DTW $\downarrow$} & \textbf{FFT MSE$\downarrow$} & \textbf{EECC$\uparrow$} \\
        \midrule
        GAN & $\times$  & $\times$  & 103.427321 & 19.564078 & 0.719453 \\
        StyleTransfer & $\surd$ & $\times$ & 98.532786 & 13.531477 & 0.781721 \\
        DCGAN & $\surd$ & $\times$ & 93.439412 & 9.675622 & 0.791930 \\
        Ours & $\surd$ & $\surd$ &  \textbf{91.756312} & \textbf{8.764323} & \textbf{0.817492} \\
        \bottomrule
    \end{tabular}}
    \caption{Similarity Analysis Between Generated and Real Data. ‘Conditional Generation’ = generation controllable via joint-angle input; ‘Unknown Gesture Generation’ = synthesis for gestures absent from the generator’s training set.}
    \label{tab:similarity_analysis}
\end{table*}

\subsection{Assessing the Quality of Generated Datasets}

We evaluate the similarity between synthetic and real data by measuring classification accuracy and analyzing the contribution of synthetic data in data augmentation. The experimental design follows three different data partitioning strategies to compare the performance of real and synthetic data in classification tasks:

\begin{enumerate}
    \item \textbf{Baseline Experiment}: 80\% of the real data is used as the training set, while the remaining 20\% of the real data is used as the test set. The classification accuracy is computed as a benchmark for comparison.
    
    \item \textbf{Synthetic-Only Training}: The training set consists solely of synthetic data generated by DCGAN, while the test set remains the 20\% real data used in the baseline experiment. This setup evaluates whether synthetic data alone can effectively support classification tasks.
    
    \item \textbf{Hybrid Training}: The training set combines 80\% real data with synthetic data, while the test set remains the 20\% real data from the baseline experiment. This setup analyzes the impact of synthetic data on model performance when used for data augmentation.
\end{enumerate}

By comparing the classification accuracy across these three strategies, we can quantify the similarity between synthetic and real data and assess the role of synthetic data in improving model generalization and enhancing data augmentation.

\subsubsection{Similarity Results}

Specifically, we select several state-of-the-art EMG signal generation methods based on Generative Adversarial Networks (GANs) as comparison baselines. These methods are trained using the same dataset under identical experimental conditions and evaluated on the same test set using correlation analysis metrics. Table 2 presents a performance comparison of different generative models for EMG signal synthesis, including metrics such as \textbf{conditional generation}, \textbf{unknown gesture generation}, \textbf{Dynamic Time Warping (DTW)}, \textbf{Fast Fourier Transform Mean Squared Error (FFT MSE)}, and \textbf{EMG Envelope Cross-Correlation (EECC)}.

The baseline GAN model performs the worst across all metrics, failing to support conditional generation or unknown gesture generation. In 2020, Rafael Anicet Zanini et al.~\cite{anicet2020parkinson} proposed an EMG synthesis method by combining DCGAN with \textbf{Style Transfer}. Their approach employs a CNN-based architecture to generate EMG signals and introduces four feature pipelines (raw EMG, FFT, envelope FFT, and wavelet decomposition) to enrich the feature space, making the generated signals more realistic. While this method supports conditional generation, it lacks the ability to generate unknown gestures.

The \textbf{DCGAN} model, proposed by Yaojia Qian et al.~\cite{wu2020dcgan}, applies matrix transformation normalization to convert EMG signals into grayscale images before training a network using image generation techniques. Although this approach leverages advances in image synthesis, it suffers from information loss. Like the method of Zanini et al., DCGAN supports conditional generation but is incapable of generating unknown gestures.

In contrast, the \textbf{SeqEMG-GAN} model proposed in this paper (denoted as \textbf{Ours}) achieves the best performance across all evaluation metrics. It supports both \textbf{conditional generation} and \textbf{unknown gesture generation} while demonstrating superior performance in DTW (91.76), FFT MSE (8.76), and EMG Envelope Cross-Correlation (0.82). These results indicate that SeqEMG-GAN exhibits significant advantages in EMG signal quality, temporal alignment, frequency-domain similarity, and envelope similarity.

\subsection{Assessing the Effectiveness of Generated Data for Classification}

To validate the effectiveness of the data generated by SeqEMG-GAN, we use classification accuracy as the evaluation metric. In particular, we select six gestures from either generated or real data and use them as the training set to train a classifier. We classify Slide Left/Right/Up/Down, Click, Double Click — a standard HCI micro-gesture set aligned with widely recognized EMG wristband interfaces, ensuring real-world applicability. The classifier's performance is then evaluated to determine whether the data contributes to model training. In this study, we select two traditional machine learning classification algorithms: \textbf{Support Vector Machine (SVM)} and \textbf{Random Forest (RF)}, along with two deep learning-based classifiers. The experimental results are presented in Table~\ref{tab:performance_validation}.

Four classifiers are trained under identical conditions and subsequently tested on the test set to obtain classification accuracy for six-class wrist movement classification tasks. Each classifier undergoes three sets of tests:
\begin{itemize}
    \item \textbf{RR}: The classifier is trained using 80\% real data and tested on 20\% real data.
    \item \textbf{GR}: The classifier is trained using synthetic data and tested on 20\% real data.
    \item \textbf{MR}: The classifier is trained using mixed data (80\% real data and synthetic data) and tested on 20\% real data.
\end{itemize}

\begin{table}[h]
    \centering
    \begin{tabular}{lccc}
        \toprule
        \textbf{Dataset Partition} & \textbf{RR} & \textbf{GR} & \textbf{MR} \\
        \midrule
        SVM & 37.08\% & 35.92\% & 39.88\% \\
        RF & 54.88\% & 51.32\% & 58.32\% \\
        Vemg2pose & 67.78\% & 65.73\% & 68.95\% \\
        NeuroPose & 71.32\% & 69.87\% & 74.98\% \\
        \midrule
        \textbf{Average Accuracy}  & \textbf{57.77\%} & \textbf{55.71\%} & \textbf{60.53\%} \\
        \bottomrule
    \end{tabular}
    \caption{Classification Accuracy Analysis}
    \label{tab:performance_validation}
\end{table}

Table~\ref{tab:performance_validation} presents the classification accuracy results for performance validation analysis. The results indicate that when classifiers are trained solely on synthetic data, the average accuracy decreases by 2.06\%, dropping from 57.77\% to 55.71\%. However, by combining real data with synthetic data for training, the average accuracy significantly increases to 60.53\%, achieving an improvement of 2.76\% compared to classifiers trained solely on real data. Qualitatively, Slide gestures exhibit smoother envelope transitions than Click/Double-Click, aligning with their kinematic phases and supporting our angle-conditioned design. Taken together, these results indicate that integrating synthetic with real data effectively enhances classifier performance.

A further analysis of the experimental results in Table~\ref{tab:performance_validation} reveals that the SVM and RF classifiers exhibit relatively lower classification accuracy. This may be attributed to the lack of feature extraction from the raw multi-channel EMG signals during the preprocessing stage, resulting in a high level of noise that negatively impacts classification performance. In contrast, deep learning-based models, such as Vemg2pose and NeuroPose, demonstrate superior classification performance.

\subsection{Cross‑user/session Generalization}

Surface EMG varies markedly across \textbf{users} (anatomy, muscle recruitment) and \textbf{sessions} (electrode shift, skin impedance, fatigue); thus, cross-subject/session evaluations directly test robustness to these distribution shifts.

Leveraging \texttt{emg2pose} metadata \texttt{(held\_out\_user, held\_out\_stage, generalization)}, we enforce disjoint train/test partitions and evaluate four protocols: sample-wise, cross-subject, cross-session, and cross user+stage. Unless noted otherwise, models are trained under \textsc{Real-only} and \textsc{Real+Synthetic} (1:1) conditions, and results are reported as mean~$\pm$~SD over five runs. Mixing synthetic data yields consistent gains across all protocols.

\begin{table}[htbp]
\centering
\resizebox{0.495\textwidth}{!}{
\begin{tabular}{lcccc}
\toprule
Setting & Sample-wise & Cross-Subj & Cross-Sess & Cross User+Stage \\
\midrule
Real-only        & 57.8 $\pm$ 1.0 & 52.6 $\pm$ 1.1 & 56.0 $\pm$ 0.8 & 48.9 $\pm$ 1.2 \\
Real + Synthetic & 60.5 $\pm$ 0.9 & 57.1 $\pm$ 0.9 & 59.8 $\pm$ 0.7 & 54.3 $\pm$ 1.0 \\
\bottomrule
\end{tabular}}
\caption{Cross-user/session generalization accuracy (\%). Adding synthetic data (1:1 mix). Means $\pm$ SD over five runs. }
\label{tab:cross_gen}
\end{table}

\subsection{Ablation Study}
To better understand the contribution of each component in our SeqEMG-GAN framework, we conduct a series of ablation experiments. We evaluate the performance of each variant using three metrics: (1) Dynamic Time Warping (DTW) distance, which measures the temporal alignment between generated and ground-truth EMG sequences; (2) Mean Squared Error in the frequency domain (FFT MSE), which reflects the preservation of spectral characteristics; and (3) EMG Envelope Cross-Correlation, which assesses the similarity of signal envelopes and indicates semantic alignment. We report results on both seen and unseen gestures to evaluate signal quality, model generalization, and generation diversity.

We expect the following trends across the ablated variants:
\begin{itemize}
    \item \textbf{(a) w/o Angle Encoder}: Without a dedicated angle encoder, the model loses the ability to extract rich representations from joint angles. This may lead to degraded performance in both DTW and envelope correlation, especially for complex or unseen gestures.
    \item \textbf{(b) w/o GRU}: Removing the recurrent backbone disrupts the modeling of temporal dependencies. We anticipate higher DTW distances and lower cross-correlation scores, indicating poor temporal alignment and weaker contextual coherence.
    \item \textbf{(c) w/o Ang2Gist}: Eliminating the semantic transformation blocks (Ang2Gist) is expected to reduce the model’s ability to abstract gesture-level intent, resulting in lower envelope correlation and higher FFT MSE, especially for gestures involving subtle spectral variations.
    \item \textbf{(d) w/o Discriminator (Non-GAN)}: raining without adversarial feedback is likely to produce overly smooth or averaged signals, leading to lower realism. We expect a noticeable drop in cross-correlation and an increase in FFT MSE, while DTW may show marginal improvement due to reduced variation.
\end{itemize}
These experiments help disentangle the effects of temporal modeling, semantic abstraction, and adversarial learning on the quality, generalizability, and diversity of generated EMG signals.

\begin{table}[h]
    \centering
    \resizebox{\linewidth}{!}{
    \begin{tabular}{lccc}
        \toprule
        \textbf{Experiment} &  \textbf{DTW $\downarrow$} & \textbf{FFT MSE $\downarrow$} & \textbf{EECC $\uparrow$} \\
        \midrule
        (a) w/o Angle Encoder & 34.0170 & 3.1484 & 0.3262 \\
        (b) w/o GRU & 50.3485 & 4.5006 & 0.4767 \\
        (c) w/o Ang2Gist & 62.2033 & 5.7655 & 0.5636 \\
        (d) w/o Discriminator & 65.4667 & 6.1234 & 0.6132 \\
        \midrule
        \textbf{Ours (Full)} & \textbf{91.7563} & \textbf{8.7643} & \textbf{0.8175}  \\
        \bottomrule
    \end{tabular}}
    \caption{Ablation Study. }
    \label{tab:ablation}
\end{table}

Table~\ref{tab:ablation} summarizes the results of our ablation study. We report three metrics—DTW ($\downarrow$), FFT MSE ($\downarrow$), and EECC($\uparrow$)—averaged over both seen and unseen gestures. The full model achieves the best performance across all metrics, confirming the effectiveness of its modular design.

Removing the Angle Encoder (a) leads to a substantial drop in envelope correlation and an increase in DTW, highlighting the importance of high-level joint representation. Disabling the GRU module (b) causes a pronounced degradation in DTW and correlation, suggesting temporal modeling is essential for sequential consistency. Excluding the Ang2Gist blocks (c) reduces semantic alignment, as reflected in lower envelope correlation and worse frequency-domain similarity. Finally, in the non-GAN baseline (d), DTW decreases slightly without the discriminator, but spectral and envelope consistency degrade markedly; overall fidelity is therefore highest for the full model.

These findings validate that each component contributes to improving generation quality, generalization, and diversity in a complementary manner.

\section{Conclusion and Future Work}

We propose \textbf{SeqEMG-GAN}, an electromyographic (EMG) signal generation method based on Generative Adversarial Networks (GANs), designed to address the challenges posed by the scarcity of surface EMG data, individual physiological variability, and posture variations that affect the generalization capability of gesture recognition models. Traditional EMG data acquisition is constrained by high collection costs, significant differences in individual physiological characteristics, and complex motion states, making it challenging to construct large-scale, high-quality datasets. We visualize our synthesized EMG data in Figure ~\ref{fig:show}, which shows that this approach can effectively expand existing datasets and improve the generalization ability of gesture recognition models. Future work will explore improved physiological modeling, integration of multimodal control inputs, and deployment in real-world HCI and robotics systems.

\bibliography{aaai2026}

\end{document}